\documentstyle[emulateapj]{article}

\submitted{to be published in the Astrophysical Journal, 519, No.1}

\lefthead{Hachisu, Kato, and Nomoto}
\righthead{Type Ia Supernovae}

\begin{document}

\title{A NEW EVOLUTIONARY PATH TO TYPE Ia SUPERNOVAE: 
HELIUM-RICH SUPER-SOFT X-RAY SOURCE CHANNEL}

\author{Izumi Hachisu}
\affil{Department of Earth Science and Astronomy, 
College of Arts and Sciences, University of Tokyo,
Komaba, Meguro-ku, Tokyo 153-8902, Japan \\ e-mail: 
hachisu@chianti.c.u-tokyo.ac.jp}

\author{Mariko Kato}
\affil{Department of Astronomy, Keio University, 
Hiyoshi, Kouhoku-ku, Yokohama 223-8521, Japan \\ e-mail: 
mariko@educ.cc.keio.ac.jp}

\author{Ken'ichi Nomoto, and Hideyuku Umeda}
\affil{Department of Astronomy, University of Tokyo, 
Bunkyo-ku, Tokyo 113-0033, Japan \\ e-mail: 
nomoto@astron.s.u-tokyo.ac.jp,
umeda@astron.s.u-tokyo.ac.jp \\
Research Center for the Early Universe, University of Tokyo, 
Bunkyo-ku, Tokyo 113-0033, Japan}




\begin{abstract}
We have found a new evolutionary path to Type Ia supernovae (SNe Ia)
which has been overlooked in previous work.  In this scenario, 
a carbon-oxygen white dwarf (C+O WD) is originated, not from an 
asymptotic giant branch star with a C+O core, 
but from a red-giant star with a helium core of $\sim 0.8-2.0 M_\odot$.
The helium star, which is formed after the first common envelope evolution,
evolves to form a C+O WD of $\sim 0.8-1.1 M_\odot$ 
with transferring a part of the helium envelope onto the secondary 
main-sequence star. 
This new evolutionary path, 
together with the optically thick wind from mass-accreting 
white dwarf, provides a much wider channel to SNe Ia 
than previous scenarios. 
A part of the progenitor systems are identified as the luminous 
supersoft X-ray sources or the recurrent novae like U Sco, which 
are characterized by the accretion of helium-rich matter.  
The white dwarf accretes hydrogen-rich, helium-enhanced 
matter from a lobe-filling, slightly evolved
companion at a critical rate and blows excess matter in the wind. 
The white dwarf grows in mass to the Chandrasekhar mass limit 
and explodes as an SN Ia.  A theoretical estimate indicates that 
this channel contributes a considerable part of the inferred rate 
of SNe Ia in our Galaxy, i.e., the rate is about ten times larger 
than the previous theoretical estimates for white dwarfs with
slightly evolved companions.  
\end{abstract}




%

\section{INTRODUCTION}
     Type Ia supernovae (SNe Ia) have been widely believed 
to be a thermonuclear explosion of a mass-accreting white dwarf (WD)
(e.g., \cite{nom97} for a recent review). 
However, the immediate progenitor binary 
systems have not been identified yet (\cite{bra95}).   
There exist two models discussed frequently as progenitors of SNe Ia:
1) the Chandrasekhar (Ch) mass model, in which a mass-accreting 
carbon-oxygen (C+O) WD
grows in mass up to the Ch mass and explodes 
as an SN Ia, and 2) the sub-Chandrasekhar (sub-Ch) mass model, 
in which an accreted layer of helium atop a C+O WD ignites 
off-center for a WD mass well below the Ch mass. 
The early time spectra of the majority of SNe Ia are 
in excellent agreement with the synthetic spectra of the Ch mass models,
while the spectra of the sub-Ch mass models are too blue to be comparable
with the observations (\cite{hof96}; \cite{nug97}).
\par
     For the evolution of accreting WDs toward the Ch mass,
two scenarios have been proposed: 
1) a double degenerate (DD) scenario, i.e., merging of double C+O WDs
with a combined mass surpassing the Ch mass limit 
(\cite{ibe84}; \cite{web84}), and 2) a single degenerate (SD) scenario,
i.e., accretion of hydrogen-rich matter via mass transfer
from a binary companion (e.g., \cite{nom82a}; \cite{nom94}).  The issue
of DD vs. SD is still debated (e.g., \cite{bra95}),
although theoretical modeling has indicated that 
the merging of WDs leads to the accretion-induced collapse 
rather than SN Ia explosion (Saio \& Nomoto 1985, 1998; \cite{seg97}).
\par
     For the Ch/SD scenario, a new evolutionary model has been 
proposed by Hachisu, Kato, \& Nomoto (1996; hereafter HKN96).
HKN96 have shown that if the accretion rate exceeds a critical rate,
the WD blows a strong wind and burns hydrogen steadily at this 
critical rate and expels excess matter in the wind.  The WD 
increases its mass up to the Ch mass avoiding formation of
a common envelope.
Li \& van den Heuvel (1997) have extended HKN96's model to 
a system consisting of a mass-accreting WD and a lobe-filling,
more massive, main-sequence (MS) or sub-giant star 
(hereafter ``WD+MS system''), identified with 
luminous supersoft X-ray sources, 
and found that such a system is one of the main progenitors 
of SNe Ia as well as a system consisting of a WD 
and a lobe-filling, less massive, red-giant (hereafter ``WD+RG system'') 
proposed by HKN96. 
\par
     Recently, Yungelson \& Livio (1998) have reanalyzed the models 
by HKN96 and Li \& van den Heuvel based on their 
population synthesis code and concluded that 
both HKN96's WD+RG and Li \& van den Heuvel's WD+MS systems 
can account for only (at most) 10\% of the inferred rate of SNe Ia 
in our Galaxy.
However, Yungelson \& Livio (1998)
overlooked important evolutionary processes 
both in the WD+MS and WD+RG systems. 
In this paper, we first describe an important 
evolutionary process to form the WD+MS system, 
which has been overlooked in previous works 
(e.g., \cite{dis94}; \cite{yun96}; \cite{yun98}). 
Another evolutionary process leading to the WD+RG system is
discussed elsewhere (Hachisu, Kato, \& Nomoto 1999, hereafter HKN99).
In \S 2, we describe the 
new evolutionary process to form the WD+MS system.  
Including this new evolutionary path, we will show that 
the secondary (slightly evolved MS star) 
becomes a helium-rich star like in U Sco 
(e.g., \cite{wil81}), which is transferring helium-rich matter 
onto the primary (WD).  We have reanalyzed such a helium-rich 
matter accretion onto the WD based on the optically thick wind 
theory developed by Kato \& Hachisu (1994).  
If the secondary MS star has a mass of $\sim 2-3.5 M_\odot$, 
a WD with an initial mass of $0.8-1.1 M_\odot$ grows in mass 
to the Ch mass and explodes as an SN Ia.  
We describe the evolution of such a WD+MS system 
in \S 3.  The new parameter region thus obtained is much wider than 
that by Li \& van den Heuvel (1997) and Yungelson \& Livio (1998).  
Discussions follow in \S 4, in which we have estimated 
the realization frequency of our WD+MS systems that accounts 
for about a third of the inferred rate of SNe Ia in our Galaxy.

\section{FORMATION OF A NAKED HELIUM CORE AND ITS EVOLUTION
TOWARD C+O WHITE DWARF}
\begin{figure}
\plotone{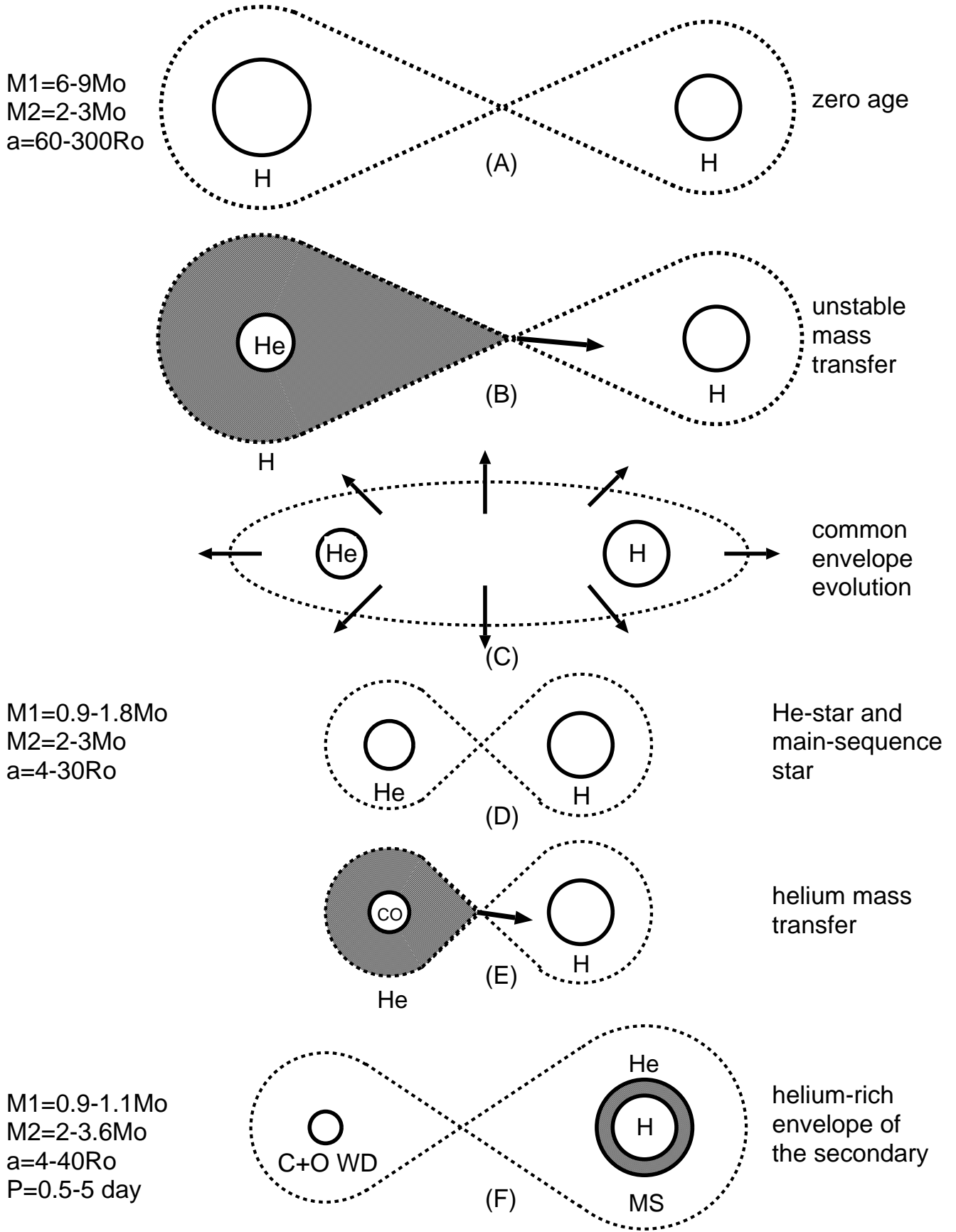}
\caption{
Early evolutionary path through the common envelope evolution to
the helium matter transfer.
\label{hemtsn1a}}
\end{figure}

      In this section, we describe the evolutionary path 
to form the WD+MS system where the secondary (MS star) has 
a helium-rich envelope.
This important evolutionary path shown in Figure \ref{hemtsn1a}
has been overlooked in previous work.
Yungelson \& Livio (1998) applied their population 
synthesis code to the WD+MS systems and found that the realization 
frequency of the WD+MS systems is at most one tenth of 
the inferred rate of SNe Ia in our Galaxy.  In their population 
synthesis code, they consider only initial systems consisting of 
a more massive AGB star with a C+O core and a less massive 
main-sequence star.  This system undergoes a common envelope evolution 
and finally yields a binary system of a mass-accreting C+O WD and 
a lobe-filling MS or sub-giant star.   Their code 
does not include another important evolutionary path, in which
a more massive component fills up its inner 
critical Roche lobe when it develops {\it a helium 
core} of $\sim 0.8-2.0 M_\odot$ in {\it its red-giant phase}.
\placefigure{hemtsn1a}

\subsection{Common Envelope Evolution at Red-Giant Phase 
with a Helium Core}
\begin{figure}
\plotone{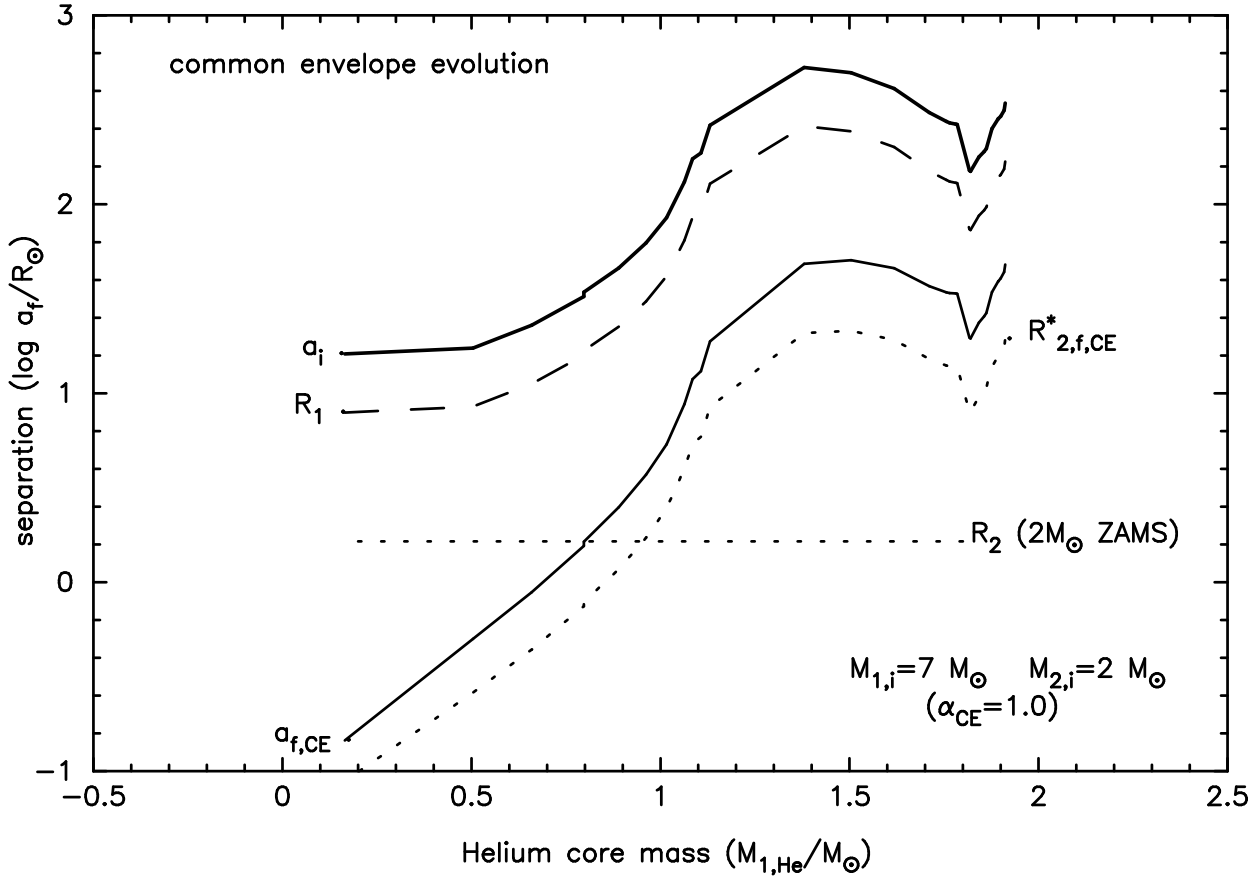}
\caption{
Long dashed line shows the radius of the primary
$R_1$ against its helium core mass $M_{\rm 1,He}$.
The radius increases with the helium core mass but it begins 
to decrease
after helium is ignited at the center of the helium core 
when $M_{\rm 1,He} \sim 1.4 M_\odot$.  
If the lobe-filling condition is satisfied, we obtain
the separation $a_i$ from equation (2).
The thick and thin solid lines show the separations 
$a_i$ (at {\it stage B} in Fig. 1) 
and $a_{f, {\rm CE}}$ (at {\it stage D}) before and after 
the common envelope evolution, respectively, 
for the $7 M_\odot + 2 M_\odot$ pair.  
The separation shrinks by about a factor of ten 
after the common envelope evolution. Dotted horizontal line 
indicates the radius of $2 M_\odot$ star 
at the zero age main-sequence (ZAMS).
The helium core mass can grow in mass for the larger 
separation $a_i$, i.e., for the larger Roche lobe $R_{1,i}^*$.
\label{m7m2com2}}
\end{figure}

     This evolutionary path from {\it stage A} to {\it F} 
in Figure \ref{hemtsn1a} has been first 
introduced by Hachisu \& Kato (1999)
to explain the helium-rich companion of the recurrent nova U Sco.
We consider, for example, a close binary at a separation $a$ 
with the primary of mass
$M_{1,i} = 7 M_\odot$ and the secondary of
$M_{2,i} = 2 M_\odot$ ({\it stage A}).  
When the primary has evolved to a red-giant of the radius $R_1$ 
forming a helium core of mass $M_{\rm 1,He}$ ({\it stage B}), 
it fills up its inner critical Roche lobe, i.e., $R_1=R_1^*$.
Here, $R^*_1$ is the effective 
radius of the inner critical Roche lobe of the primary,
which is approximated by Eggleton's (1983) formula,
\begin{equation}
{{R_1^*} \over {a}} = f(q) \equiv  {{0.49q^{2/3}}
\over{0.6q^{2/3}   +  \ln  (1+q^{1/3})}},
\label{inner_critical_Roche_lobe}
\end{equation}
for the mass ratio $q= M_1/M_2$.
The long dashed line in Figure \ref{m7m2com2} shows 
the primary radius $R_1$ 
against the helium core mass $M_{\rm 1,He}$ (\cite{bre93}).
The radius of the primary increases with the helium core mass 
from $M_{\rm 1,He} \sim 0.2 M_\odot$ 
to $\sim 1.4 M_\odot$ until
helium burning ignites at the center of the helium star.  
If the primary fills its inner critical Roche lobe 
at a certain $M_{\rm 1,He}$, i.e., $R_1(M_{\rm 1,He}) = R_1^*$, 
the separation of the binary, $a$, is given by 
\begin{equation}
a = {{R_1(M_{\rm 1,He})} \over {f(q)}}.  
\label{inner_separation}
\end{equation}
Using this relation, we can plot the initial separation $a_i$  
against $M_{\rm 1,He}$ as shown by the thick solid line 
in Figure \ref{m7m2com2}.
\par
     Then the mass transfer begins.  
This mass transfer is dynamically unstable 
because the primary red-giant star has a convective envelope. 
The binary undergoes common envelope (CE) evolution 
({\it stage C}), which yields a much more compact close binary 
consisting of a naked helium star of $M_{1,{\rm He}}$ and 
a main-sequence star of $M_2= 2 M_\odot$ ({\it stage D}). 
Figure \ref{m7m2com2} shows the separation $a_{f,{\rm CE}}$ and
the inner critical Roche lobe radius 
of the secondary $R^*_{2,f,{\rm CE}}$ after the common envelope
evolution.  Here, we assume the relation
\begin{equation}
{{a_{f,{\rm CE}}} \over {a_{i}}} \sim \alpha_{\rm CE} \cdot 
\left({{M_{1,{\rm He}}} \over { M_{1,i}}} \right) 
\cdot \left({{M_2} \over {M_{1,i}-M_{1,{\rm He}}}} \right),
\label{common_envelope_separation}
\end{equation}
with the efficiency $\alpha_{\rm CE}=1.0$ 
for the common envelope evolution
(e.g., \cite{ibe84}; \cite{ibe93}; \cite{yun98}).
\par
     After the common envelope evolution, 
the radii of the inner critical Roche lobes of the primary and 
the secondary become $R_1^* \sim 0.36 a_{f, \rm{CE}}$ and 
$R_2^* \sim 0.4 a_{f, \rm{CE}}$, respectively.
Since the secondary radius (shown by the dashed line 
in Figure \ref{m7m2com2} for the $2 M_\odot$ ZAMS) 
should be smaller than 
its inner critical Roche lobe, i.e., 
$R_2 < R^*_{2,f,{\rm CE}}$, 
the initial separation $a_{i}$ should 
exceed $\sim 80 ~R_\odot$ as shown in Figure \ref{m7m2com2}.  
The upper bound of the initial separation 
is obtained from the maximum radius of the $7 M_\odot$ star 
which has formed a helium core, i.e., 
$a_{i} \approx 2 R_{\rm 1,max} \lesssim 2 \times 300 ~R_\odot$. 
Thus, the allowable range of the initial separations for our model 
is $80 ~R_\odot \lesssim a_{i} \lesssim 600 ~R_\odot$, so that 
$R_1^* \sim 1.4-18 ~R_\odot$ and $R_2^* \sim 1.6-20 ~R_\odot$
for the case of Figure \ref{m7m2com2}. 

\placefigure{m7m2com2}

\subsection{Transfer of Helium Envelope to the Secondary}
\begin{figure}
\plotone{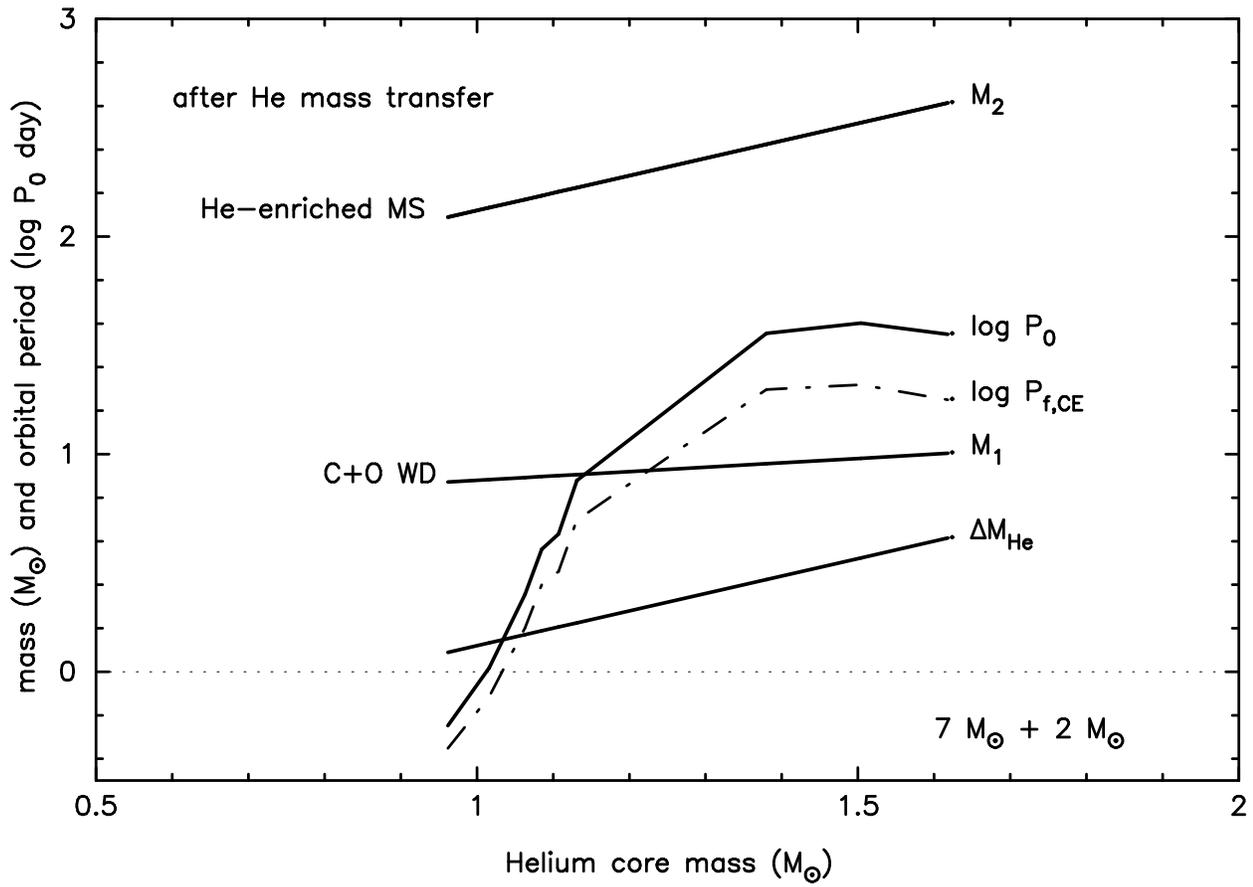}
\caption{
Thick solid lines show
the mass of the primary (C+O WD) $M_1$, the mass of the secondary
(helium-enriched MS) $M_2$, 
the mass transferred to the secondary $\Delta M_{\rm He}$,
and the orbital period $\log P_0$ after the helium mass transfer 
({\it stage F})
against the helium core mass at the beginning of helium mass 
transfer ({\it stage E}).  
The Dash-dotted line indicates the orbital period at the beginning
of helium mass transfer ({\it stage E}).  
The separation increases and the orbital
period also increases after the helium mass transfer. 
\label{m7m2hemt}}
\end{figure}

     After the hydrogen-rich envelope is stripped away and 
hydrogen shell burning vanishes, the naked helium core 
contracts to ignite central helium burning and becomes 
a helium main sequence star of mass $M_{\rm 1,He}$ ({\it stage D}).  
For $M_{\rm 1,He} \lesssim 2 M_\odot$, its C+O core mass is less
than $1.07 M_\odot$ which is the lower mass limit to 
the non-degenerate carbon ignition 
(e.g., \cite{ume99}; \cite{nom88} for a review).  
Then the helium star forms a degenerate C+O core,
whose mass $M_{\rm C+O}$ grows by helium shell burning. 
When $M_{\rm C+O}$ becomes $0.9-1.0 M_\odot$ 
and the core becomes strongly degenerate, its helium envelope 
expands to $R_1 \sim 1.4-18 ~R_\odot$ 
(e.g., \cite{pac71a}; \cite{nom82b}) to 
fill its inner critical Roche lobe again ({\it stage E}).  
Helium is 
transferred to the secondary stably on an {\it evolutionary 
time scale} of $\tau_{\rm EV} \sim 10^5$ yr because 
the mass ratio is smaller than $0.79$ ($q= M_1/M_2 < 0.79$).
The resultant mass transfer (MT) rate is 
$\dot M_1 \sim 10^{-5} M_\odot$ yr$^{-1}$, which is too low  
to form a common envelope (e.g., \cite{neo77}; \cite{kip77}). 
After the helium envelope is lost, the primary becomes a C+O WD 
of $M_{\rm WD} \sim 0.9-1.1 M_\odot$ ({\it stage F}).
In Figure \ref{m7m2hemt}, $M_{1,f,{\rm MT}}= M_{\rm WD}$ is 
plotted as a function of $M_{\rm 1,He}$. 
Here, we assume the relation of (in solar mass units of $M_\odot$)
\begin{equation}
M_{\rm WD} = \left\{ 
      \begin{array}{@{\,}ll}
        0.2 \left( M_{\rm 1,He} - 0.85 \right)+0.85,
& \mbox{~for~} 0.85 < M_{\rm 1,He} \lesssim 2, \cr
M_{\rm 1,He}, 
& \mbox{~for~} 0.46 \lesssim M_{\rm 1,He} \le 0.85,
\end{array}
\right.
\label{C+O_core_mass}
\end{equation}
for the final degenerate C+O WD mass 
vs. the initial helium star mass relation
(reduced from the evolutionary paths by \cite{pac71a} 
and \cite{nom82b}). 
Helium stars with $M_{\rm He} \le 0.85 M_\odot$ do not
expand to $\sim 2-10 R_\odot$, thus burning most of helium to C+O  
without transferring helium to the secondary.  
For $M_{\rm 1,He} \gtrsim 2.0 M_\odot$, the C+O core has a mass 
larger than $1.07 M_\odot$ before becoming degenerate, thus igniting
carbon to form an O+Ne+Mg core
(e.g., \cite{nom84}), thus we do not include their mass range.
For $M_{\rm 1,He} \lesssim 0.46 M_\odot$, helium is not ignited to 
form a C+O core.
\par  
     The secondary increases its mass $M_2$ by receiving 
almost pure helium matter of 
$\Delta M_{\rm He} \sim 0.1-0.6 M_\odot$, which is 
a function of $M_{\rm 1,He}$ as plotted in Figure \ref{m7m2hemt}. 
Then a helium-enriched envelope is formed as illustrated 
in Figure \ref{hemtsn1a} ({\it stage F}).
After the helium mass transfer (MT), 
the separation increases by 10\%$-$40\%, i.e., 
$a_{\rm f, MT} \sim (1.1-1.4)
a_{\rm f, CE} \sim (4-70) R_\odot$. 
Here, we assume the conservation of the total mass and angular momentum 
during the helium mass transfer, which leads to the relation of
\begin{equation}
{{a_{\rm f, MT}} \over {a_{\rm f, CE}}} = \left({{M_{\rm1,He}} \over
{{M_{\rm 1,He}-\Delta M_{\rm He}}}} \right)^2
\left({{M_{2,i}} \over {M_{2,i} + \Delta M_{\rm He}}} \right)^2,
\label{helium_mass_transfer}
\end{equation}
and we use this to obtain $a_{\rm f, MT}$ and then the orbital 
period $P_0 \equiv P_{\rm f, MT}$ in Figure \ref{m7m2hemt}.
\placefigure{m7m2hemt}
\par
     Since the secondary receives $\sim 0.1-0.6 M_\odot$ helium, 
its hydrogen content in the envelope decreases to 
$X \sim 0.6$ if helium is completely mixed into the central 
part of the star.  However, the envelope of 
the mass-receiving star is not convective but radiative so that 
the helium content may be higher in the outer part of the star.

\subsection{Helium-Enriched Main-Sequence Companion}
     We have examined total $5 \times 5= 25$ cases, 
$M_{1,i}= 4$, 5, 6, 7, $9 M_\odot$ and $M_{2,i}= 1.0$, 1.5, 2.0, 2.5, 
$3.0 M_\odot$, and have found the possible progenitors to be 
in the range of 
$M_{\rm 1, C+O ~WD} \sim 0.8-1.1 M_\odot$ and 
$M_{\rm 2, MS} \sim 1.7-3.5 M_\odot$ with the separation of 
$a_{\rm f, MT} \sim 4-80 ~R_\odot$. 
In these cases, the secondary forms a helium-enriched envelope
for the primary mass of $M_{\rm 1, WD} \sim 0.9-1.1 M_\odot$
(but not for $M_{\rm 1, WD} \sim 0.8-0.85 M_\odot$). 
We assume, in this paper, 
that the average mass fractions of hydrogen and helium 
in the envelope are $X= 0.50$ and $Y= 0.48$ respectively.
(assuming the solar abundance of heavy elements $Z= 0.02$).
For the $9 M_\odot + 2.5 M_\odot$ case, much more helium is transferred,
while much less helium is transferred 
for the $6 M_\odot + 2 M_\odot$ case (see Table \ref{tbl-1}).

\section{GROWTH OF C+O WHITE DWARFS}
\begin{figure}
\plotone{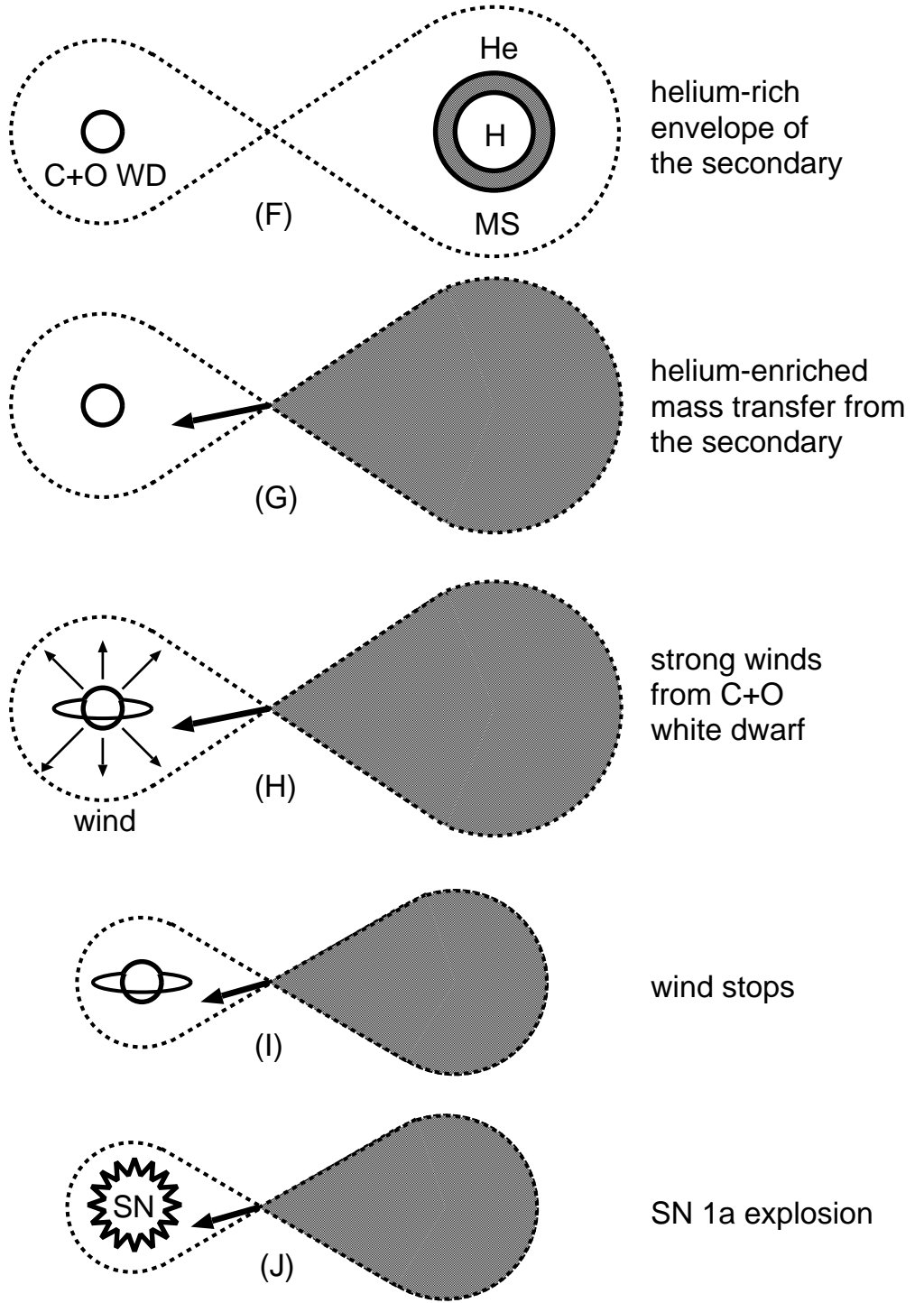}
\caption{
Late evolutionary path to an SN Ia explosion in our wind model.
\label{windsn1a}}
\end{figure}

     Starting from a close binary of 
$M_{\rm WD, 0} \equiv M_{\rm 1, C+O ~WD} \sim 0.8-1.1 M_\odot$ and 
$M_{\rm MS, 0} \equiv M_{\rm 2, MS} \sim 1.7-3.5 M_\odot$ 
with the separation of 
$a_0 \equiv a_{\rm f, MT} \sim 4-80 ~R_\odot$, 
we have followed a growth of the WD component to examine whether
the WD reaches $1.38 M_\odot$ and explodes as an SN Ia
(from {\it stage F} to {\it J} in Fig. \ref{windsn1a}).
\par
     The initial secondary now has a helium-rich envelope 
({\it stage F}).  It evolves 
to expand and fills its inner critical Roche lobe 
near the end of main-sequence (MS) phase.  Then, it starts mass transfer
({\it stage G}).  This is a {\it case A mass transfer} 
after Kippenhahn \& Weigert (1967, also \cite{pac71b}) or a
{\it cataclysmic-like mass transfer} after Iben \& Tutukov (1984). 
Since the donor is more massive than the accretor (WD component), 
the separation decreases and the inner critical Roche lobe 
decreases even to scrape the envelope mass off the donor star.  
Thus, the 
mass transfer proceeds on a {\it thermal time scale} rather 
than an evolutionary time scale ({\it stage H}).
The transferred matter is helium-rich as observed in the recurrent nova 
U Sco.  
\placefigure{windsn1a}

\subsection{Optically Thick Winds from Mass-Accreting White Dwarfs}
\begin{figure}
\plotone{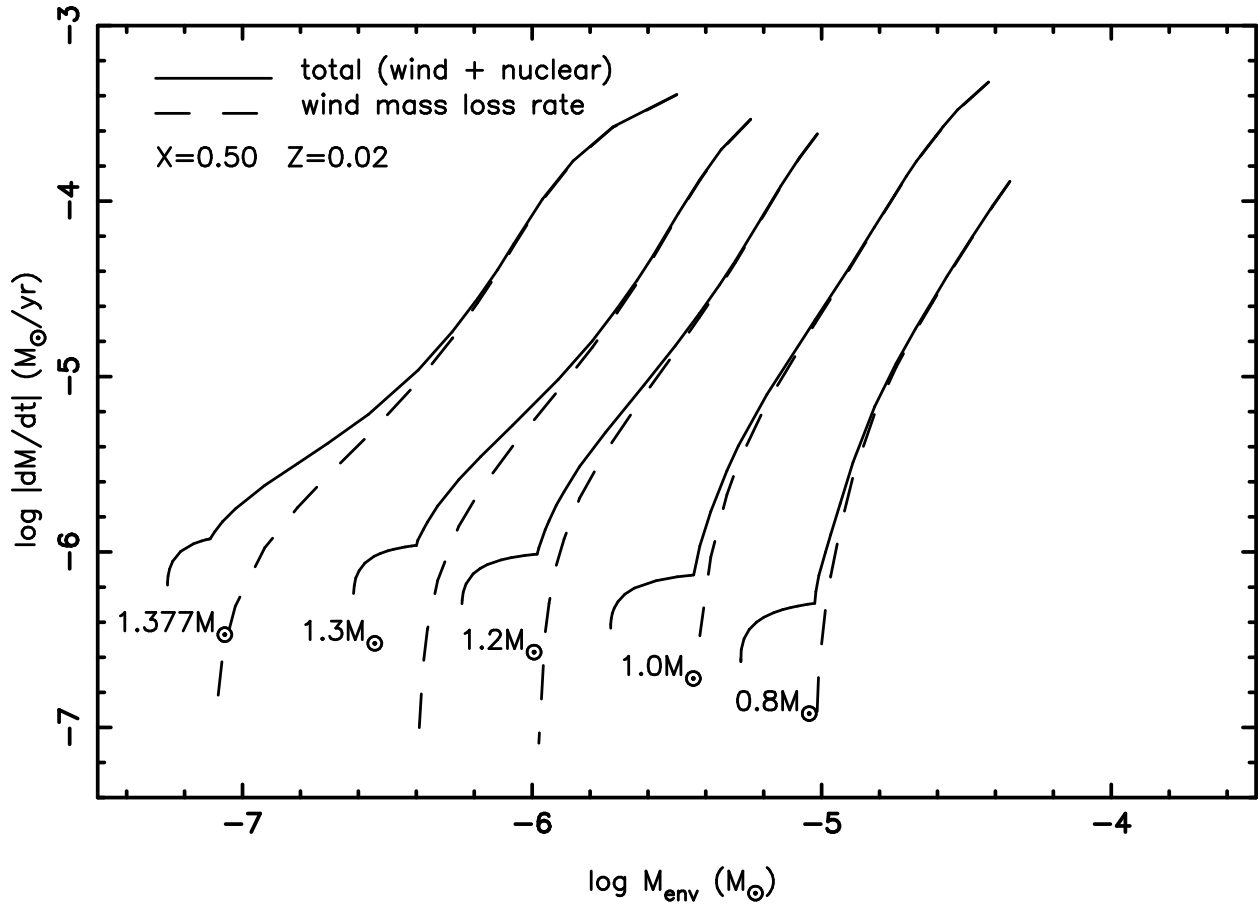}
\caption{
Wind mass loss rate (dashed line) and envelope mass
decreasing rate (solid line), i.e., nuclear burning rate plus 
wind mass loss
rate, are plotted against the envelope mass for white dwarfs with 
mass of 0.8, 1.0, 1.2, 1.3 and $ 1.377 M_\odot$.  The white dwarf 
mass is attached to each line.  There exist only static solutions
below the break on each solid line while optically thick winds blow 
above the break. 
\label{dmdtenvx50z02}}
\end{figure}

\begin{figure}
\plotone{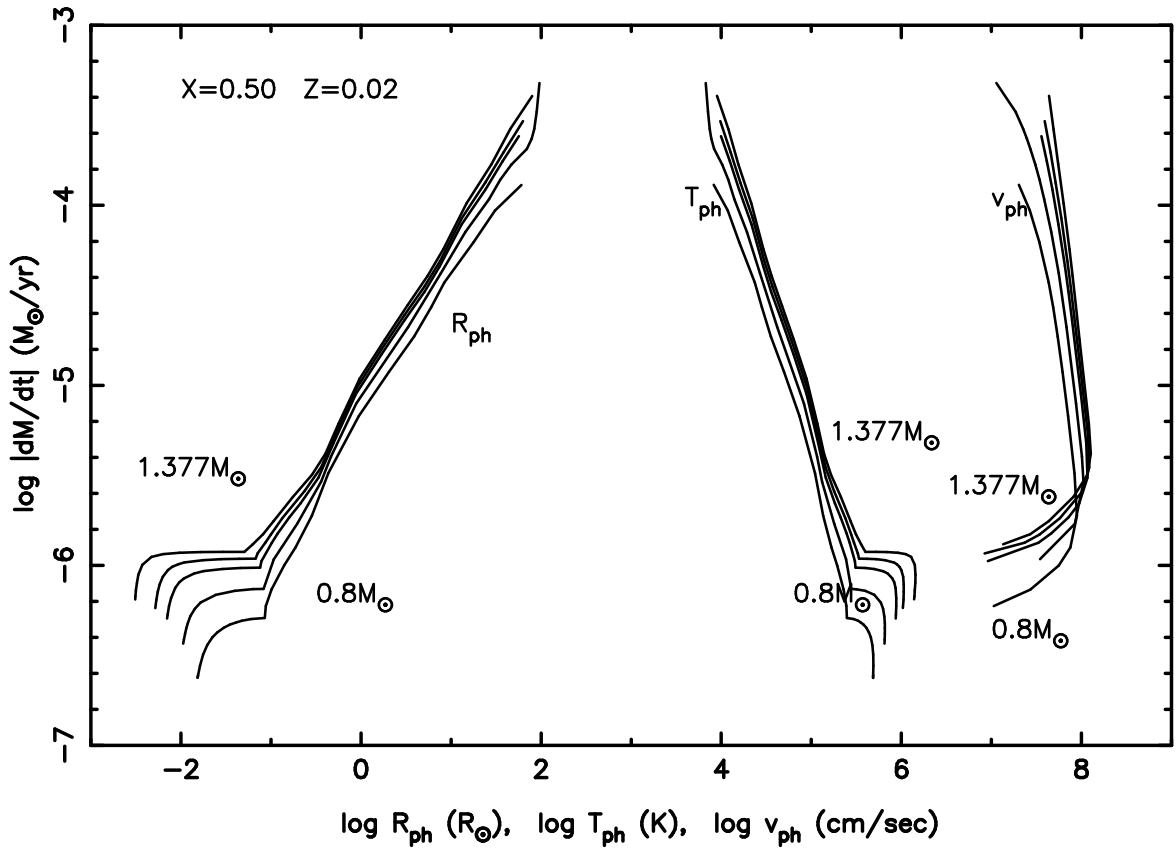}
\caption{
Photospheric radius ($R_{\rm ph}$), temperature ($T_{\rm ph}$),
and velocity ($v_{\rm ph}$) are plotted against the envelope mass
decreasing rate (i.e., nuclear burning rate plus 
wind mass loss rate) for five cases of the white dwarf mass,
$M_{\rm WD}=0.8$, 1.0, 1.2, 1.3 and $ 1.377 M_\odot$.
There exist only static solutions
below the break on each solid line of the photospheric 
temperature and radius while optically thick winds blow 
above the break.  Photospheric velocity is plotted only for
wind solutions. 
\label{dmdttrvx50z02}}
\end{figure}

     Hachisu, Kato, \& Nomoto (1996, HKN96) have shown 
that optically thick winds blow from the white dwarf
when the mass accretion rate exceeds a critical value ({\it stage H}).
In the present case, the accreted matter to form the white dwarf 
envelope is helium-rich, which is different from 
the solar abundance in HKN96.  Assuming 
$X= 0.50$, $Y= 0.48$, and $Z= 0.02$,
we have calculated the envelope models of accreting white dwarfs
for various accretion rates and white dwarf masses, 
i.e., $M_{\rm WD}= 0.6$, $0.7$, $0.8$, $0.9$, $1.0$, $1.1$, $1.2$, $1.3$, 
$1.35$, and $1.377 M_\odot$.   Optically thick winds occur 
for all these ten cases of $M_{\rm WD}$ as shown for five cases of
$M_{\rm WD}= 0.8$, $1.0$, $1.2$, $1.3$, and $1.377 M_\odot$ 
in Figures \ref{dmdtenvx50z02} and \ref{dmdttrvx50z02}.
Our numerical methods have been described in Kato \& Hachisu (1994).  
The envelope solution is uniquely determined if the envelope mass
$M_{\rm env}$ is given, where $M_{\rm env}$ is the
mass above the base of the hydrogen-burning shell.  
Therefore, the wind mass loss rate 
$\dot M_{\rm wind}$ and the nuclear burning rate $\dot M_{\rm nuc}$
are obtained as a function of the envelope mass $M_{\rm env}$, i.e.,  
$\dot M_{\rm wind} (M_{\rm env})$ and $\dot M_{\rm nuc} (M_{\rm env})$
as shown in Figure \ref{dmdtenvx50z02}.
\par
     The envelope mass of the white dwarf is determined by
\begin{equation}
\dot M_{\rm env} = \dot M_2 - (\dot M_{\rm wind}+ \dot M_{\rm nuc}).
\label{envelope_mass_rate}
\end{equation}
If the mass transfer rate does not change 
much in a thermal time scale of the WD envelope,
the steady-state $\dot M_{\rm env}=0$, i.e., 
\begin{equation}
\dot M_2 = \dot M_{\rm wind}+\dot M_{\rm nuc},
\label{envelope_mass_equilibrium}
\end{equation}
is a good approximation.   In such a steady-state approximation,
the ordinates in Figures \ref{dmdtenvx50z02} and \ref{dmdttrvx50z02},
$\dot M_{\rm wind} + \dot M_{\rm nuc}$,
are regarded as the mass transfer rate from the secondary
$\dot M_2$.  
Thus, the envelope solution is determined from the relation 
in Figure \ref{dmdtenvx50z02} for the given mass transfer 
rate $\dot M_2$.  The photospheric radius, 
temperature, and velocity are also obtained from 
the relations in Figure \ref{dmdttrvx50z02}.
\par
     Optically thick winds blow when the mass transfer rate exceeds
the critical rate, which corresponds to the break of each solid line
in Figure \ref{dmdtenvx50z02}.
There exists only a static (no wind) envelope solution 
for the mass transfer rate below this break.  
Its critical accretion rate is approximated as 
\begin{equation}
\dot M_{\rm cr} = 1.2 \times10^{-6} 
\left({M_{\rm WD} \over {M_\odot}}  -
0.40\right) M_\odot {\rm ~yr}^{-1}, 
\label{critical}
\end{equation}
for $X=0.50$, $Y=0.48$ and $Z=0.02$. 
If the mass accretion rate exceeds the critical rate, i.e.,
$\dot M_2 > \dot M_{\rm cr}$,
the strong wind blows from the white dwarf.
The white dwarf accretes helium almost at the critical rate, 
i.e., $\dot M_{\rm nuc} \approx \dot M_{\rm cr}$, 
and expels the excess matter in the wind at a rate of
$\dot M_{\rm wind} \approx \dot M_{2} - \dot M_{\rm cr}$.
Here, we assume that the white dwarf accretes 
helium-rich matter from the equator via 
the accretion disk and blows winds to the pole or off the equator.
\placefigure{dmdtenvx50z02}
\placefigure{dmdttrvx50z02}

\subsection{Efficiency of Mass-Accretion in Hydrogen Shell Burning}
     Steady hydrogen shell burning converts hydrogen into
helium atop the C+O core, which can be regarded as 
a helium matter accretion onto the C+O WD.
To estimate whether or not the white dwarf mass grows 
to $1.38 M_\odot$, 
we must calculate the mass accumulation efficiency, that is, 
the ratio between the mass accumulated in the WD after 
H/He-burning and the mass transferred from the companion. 
We denote the efficiency by $\eta_{\rm H}$ and $\eta_{\rm He}$ 
for hydrogen shell-burning and helium shell-burning, respectively. 
As shown in the previous subsection, the excess matter is blown 
in the wind when the mass transfer rate exceeds the critical rate,
which leads to  
$\eta_{\rm H}= (\dot M_2 - \dot M_{\rm wind})/\dot M_2 < 1$,
for the wind phase.
\par
     During the evolution of mass-accreting white dwarfs,
the accretion rate becomes lower than $\dot M_{\rm cr}$ in some cases.  
Then the wind stops ({\it stage I} in Fig. \ref{windsn1a}).
Hydrogen steadily burns for $\dot M_2 > 
\dot M_{\rm st} \approx 0.5 \dot M_{\rm cr}$.  
Then, we have $\eta_{\rm H}=1$.
For $\dot M_2 < \dot M_{\rm st}$, 
hydrogen shell-burning becomes unstable 
to trigger weak shell flashes.  
Once a weak hydrogen shell flash occurs, a part of the envelope mass 
of the white dwarf may be lost from the system (e.g., \cite{kov94}).   
In the present study, 
no mass loss is assumed during the weak hydrogen shell flashes 
until the mass transfer rate becomes lower than
$\dot M_{\rm low} = 1 \times 10^{-7} M_\odot \mbox{~yr}^{-1}$, i.e.,
$\eta_{\rm H}=1$ 
for $\dot M_{\rm low} \lesssim \dot M_2 < \dot M_{\rm st}$.  
When the mass transfer rate 
becomes lower than $\dot M_{\rm low}$, however,
no mass accumulation is expected from such relatively strong 
hydrogen shell-flashes (e.g., \cite{kov94}), 
i.e., $\eta_{\rm H}=0$ for 
$\dot M_2 \lesssim \dot M_{\rm low}$.   
Therefore, we stop calculating the binary evolution either when 
the primary reaches $1.38 M_\odot$
({\it stage J} in Fig. \ref{windsn1a}), i.e., 
$M_{\rm 1, WD}= 1.38 M_\odot$ or the mass transfer rate becomes
lower than $\dot M_{\rm low}$.
\par
     To summarize, we have used the following simplified relation
\begin{equation}
\eta_{\rm H}= \left\{ 
      \begin{array}{@{\,}ll}
        (\dot M_2 - \dot M_{\rm wind})/ \dot M_2 < 1,
&(\dot M_{\rm cr} < \dot M_2 < 1 \times 10^{-4} M_\odot 
\mbox{~yr~}^{-1}) \cr
1, 
& (\dot M_{\rm low} \le \dot M_2 \le \dot M_{\rm cr}) \cr
0, 
& (\dot M_2 \le \dot M_{\rm low}) \cr
\end{array}
\right.
\label{hydrogen_accumulation}
\end{equation}
for the mass accumulation efficiency of hydrogen shell burning.

\subsection{Efficiency of Mass-Accretion in Helium Shell Burning}
     For $\dot M_2 \ge \dot M_{\rm cr}$, steady hydrogen burning 
is equivalent to the helium accretion at the critical rate of 
$\dot M_{\rm cr}$ given by equation (\ref{critical}).
In this case,
weak helium shell flashes are triggered 
(e.g., \cite{kat89}) and almost all processed matter is 
accumulated on the C+O WD.  
Recently, Kato \& Hachisu (1999) have recalculated 
the helium wind model after the helium shell flashes and estimated 
the mass accumulation efficiency with the updated OPAL opacity
(\cite{igl96}).  
Here, we adopt their new results in a simple analytic form, i.e., 
\begin{equation}
\eta_{\rm He} = \left\{ 
      \begin{array}{@{\,}ll}
1,  
& (-5.9 \le \log \dot M_{\rm He} \lesssim -5) \cr
-0.175 ( \log \dot M_{\rm He} & + ~~5.35 )^2 + 1.05, \cr
& (-7.8 < \log \dot M < -5.9) 
       \end{array}
    \right.
\label{helium_accumulation}
\end{equation}
where the helium mass accretion rate, $\dot M_{\rm He}$, 
is in units of $M_\odot$ yr$^{-1}$.  We use this formula 
for various white dwarf masses and accretion rates,
although their results are given only for 
the $1.3 M_\odot$ white dwarf (\cite{kat99h}).
\par
     The wind velocity in helium shell flashes reaches 
as high as $\sim 1000$ km s$^{-1}$ (\cite{kat99h}),
which is much faster than the orbital velocities of our
WD+MS binary systems $a\Omega_{\rm orb} \sim 300$ km s$^{-1}$.
It should be noted that either a Roche lobe overflow or a  
common envelope does not play a role as a mass ejection mechanism
because the envelope matter goes away quickly from the system 
without interacting with the orbital motion 
(see \cite{kat99h} for more details).

\subsection{Mass Transfer Rate of the Secondary}
     We have followed binary evolutions from the initial 
state of $(M_{\rm WD,0}, M_{\rm MS,0}, a_0)$ 
or $(M_{\rm WD,0}, M_{\rm MS,0}, P_0)$,
where $P_0$ is the initial orbital period.  
Here, the subscript naught (0) denotes {\it stage F} in
Figure \ref{windsn1a}, that is, before the mass transfer 
from the secondary starts.  The radius and luminosity 
of slightly evolved main-sequence stars are calculated 
from the analytic form given by Tout et al. (1997). 
The mass transfer proceeds on a thermal time scale 
for the mass ratio of $M_2/M_1 > 0.79$.  We approximate 
the mass transfer rate as
\begin{equation}
\dot M_2  = {{M_2} \over {\tau_{\rm KH}}} \cdot \max\left(
{{\zeta_{\rm RL} - \zeta_{\rm MS}} \over {\zeta_{\rm MS}}},0 \right),
\label{secondary_mass_transfer}
\end{equation}
where $\tau_{\rm KH}$ is the Kelvin-Helmholtz timescale (e.g., 
\cite{pac71b}), and $\zeta_{\rm RL}=d \log R^*/d \log M$ and 
$\zeta_{\rm MS}= d \log R_{\rm MS}/d \log M$ are 
the mass-radius exponents of the inner critical Roche lobe and 
the main sequence component, respectively (e.g., \cite{hje87}).
The effective radius of the inner critical Roche lobe, $R^*$, is
calculated from equation (\ref{inner_critical_Roche_lobe}).

\subsection{Late Binary Evolution toward SN Ia}
     The separation is determined by
\begin{equation}
{{\dot a} \over {a}} = {{\dot M_1 + \dot M_2} \over {M_1+M_2}}
- 2 {{\dot M_1} \over {M_1}} -2 {{\dot M_2} \over {M_2}}
+ 2 {{\dot J} \over {J}}.
\label{separation_change}
\end{equation}
We estimate the total mass and angular momentum losses by the winds as
\begin{equation}
\dot M \equiv \dot M_1 + \dot M_2 = \dot M_{\rm wind},
\label{wind_mass_loss}
\end{equation}
and
\begin{equation}
{{\dot J} \over {J}} 
= \ell \cdot {{(M_1+M_2)^2} \over {M_1 M_2}} {{\dot M} \over {M}},
\label{specific_angular_momentum_loss}
\end{equation}
where $\ell$ is a numeric factor expressing 
the specific angular momentum of the wind, i.e.,
\begin{equation}
{{\dot J} \over {\dot M}} = \ell \cdot a^2 \Omega_{\rm orb}.
\label{specific_angular_momentum_wind}
\end{equation}
For the very fast winds such as 
$v_{\rm wind} \gtrsim 2 a \Omega_{\rm orb}$, 
the wind has, on average, the same specific angular momentum 
as that of the WD component, i.e., 
\begin{equation}
\ell = \left( {{M_2} \over {M_1 + M_2}} \right)^2,
\label{ell_value}
\end{equation}
because the wind is too fast to interact with the orbital motion
(see also HKN99).
In equations 
(\ref{separation_change})$-$(\ref{specific_angular_momentum_wind}),
we must take into account the sign of the mass loss rates, i.e.,
$\dot M = \dot M_{\rm wind} \le 0$, $\dot M_2 \le 0$, $\dot J \le 0$, 
and so on. 
\placefigure{evlx50z02}
\par
     Figure \ref{evlx50z02} shows an example of such close 
binary evolutions which lead to the SN Ia explosion.  Here,
we start the calculation when the secondary fills its inner 
critical Roche lobe.  The initial parameters are
$M_{\rm WD,0}= 1.0 M_\odot$, $M_{\rm MS,0}= 2.0 M_\odot$, 
and $a_0= 9.6 R_\odot$ ($P_0= 2.0$ d).  
The mass transfer begins at a rate as high as
$\dot M_{\rm 2} = 2.2 \times 10^{-6} M_\odot {\rm ~yr}^{-1}$.
The WD burns hydrogen to form a helium layer at the critical rate of 
$\dot M_{\rm cr} = 0.7 \times 10^{-6} M_\odot {\rm ~yr}^{-1}$ and
the wind mass loss rate is 
$\dot M_{\rm wind} = 1.5 \times 10^{-6} M_\odot {\rm ~yr}^{-1}$.
Thus a large part of the transferred 
matter is blown off in the wind.  
\par
     Since the mass ratio $M_2/M_1$ decreases,  
the mass transfer rate determined 
by equation(\ref{secondary_mass_transfer})
gradually decreases below $\dot M_{\rm cr}$.  
The wind stops at $t= 1.9 \times 10^5 {\rm ~yr}$. 
The mass transfer rate becomes lower than
$\dot M_{\rm st} \sim 5 \times 10^{-7} M_\odot$ yr$^{-1}$ (for 
$M_{\rm WD}= 1.2 M_\odot$) 
at $t= 3.8 \times 10^5 {\rm ~yr}$ and very weak shell flashes 
may occur.  
The WD mass gradually grows to reach $1.38 M_\odot$
at $t= 6.7 \times 10^5$ yr.  At this time,
the mass transfer rate is still as high as 
$\dot M_2 = 3.6 \times 10^{-7} M_\odot {\rm ~yr}^{-1}$, because 
the mass ratio $M_2/M_1$ is still larger than $0.79$, implying 
thermally unstable mass transfer.
\par
     This WD+MS system may not be observed in X-rays 
during the strong wind phase 
due to the self-absorption of X-rays. However, it is certainly 
identified as a luminous supersoft X-ray source
from $t= 1.9 \times 10^5 {\rm ~yr}$ to $3.8 \times 10^5 {\rm ~yr}$,
because it is in a steady hydrogen shell burning phase without 
a strong wind.  Just before the explosion, it may be observed 
as a recurrent nova like U Sco, which indicates 
a helium-rich accretion in quiescence (\cite{han85}).

\subsection{Outcome of Late Binary Evolution}
\begin{figure}
\plotone{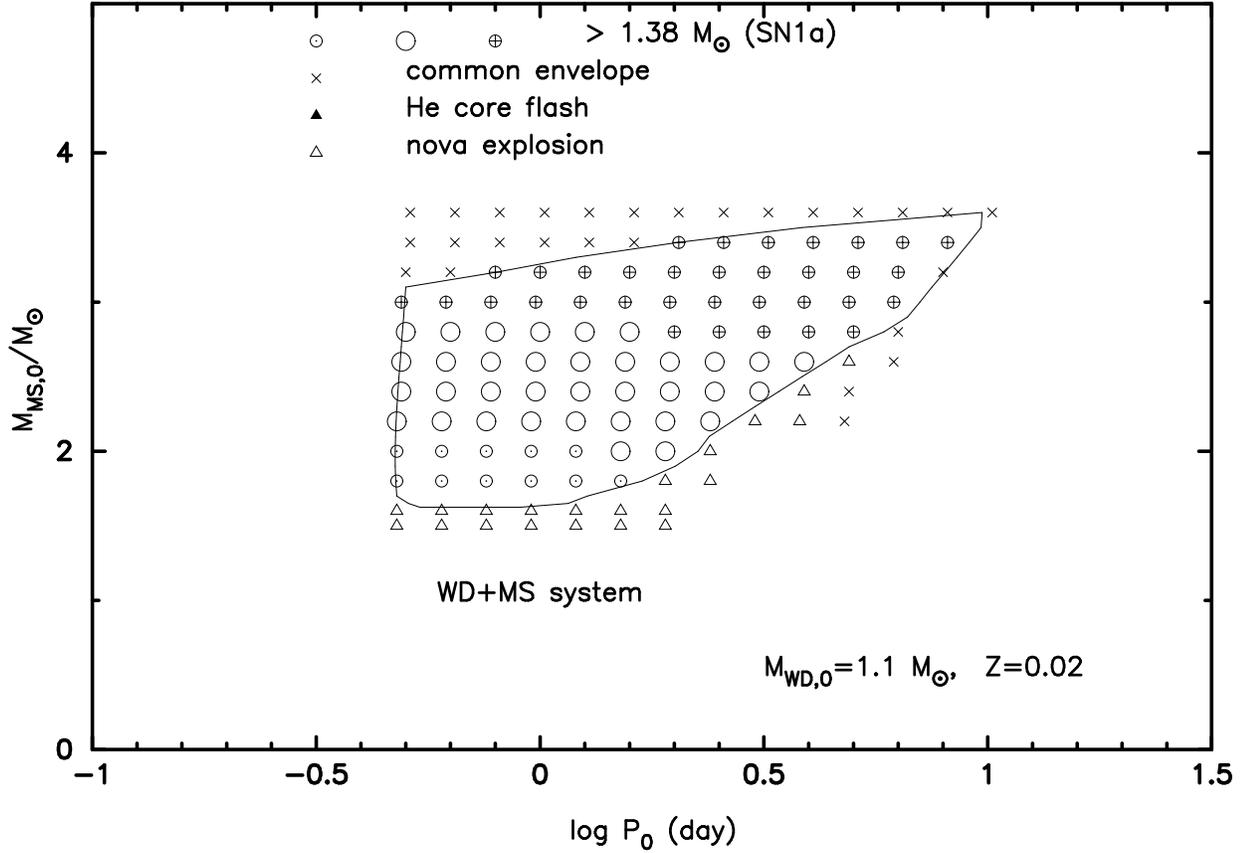}
\caption{
Final outcome of close binary evolution 
in the $\log P_0 - M_{\rm MS,0}$ plane for $M_{\rm WD,0}= 1.1 M_\odot$.  
Final outcome is either 
1) forming a common envelope (denoted by $\times$) because 
the mass transfer rate at the beginning is large enough to form 
a common envelope, i.e., $R_{\rm ph} \gtrsim 10 R_\odot$ for 
$\dot M_2 \gtrsim 1 \times 10^{-4}$ yr$^{-1}$ as seen in Fig. 6, 
2) triggering an SN Ia explosion 
(denoted by $\oplus$, $\bigcirc$, or $\odot$) 
when $M_{\rm 1,WD}= 1.38 M_\odot$
or 3) triggering repeated nova cycles (denoted $\triangle$), i.e.,  
$\dot M_2 < \dot M_{\rm low}$ when $M_{\rm 1,WD} < 1.38 M_\odot$.
Among the SN Ia cases, the wind status at the explosion depends on 
$\dot M_2$ as follows.  2a) Wind continues at the SN Ia explosion
for $ \dot M_{\rm cr} <  \dot M_2 \lesssim 1 \times 10^{-4} M_\odot$ 
yr$^{-1}$  ($\oplus$).
2b) Wind stops 
before the SN Ia explosion but the mass transfer rate is still 
high enough to keep steady hydrogen shell burning for
$ \dot M_{\rm st} <  \dot M_2 < \dot M_{\rm cr}$ ($\bigcirc$). 
2c) Wind stops before the SN Ia explosion and the mass transfer 
rate decreases to between 
$ \dot M_{\rm low} <  \dot M_2 < \dot M_{\rm st}$ 
at the SN Ia explosion ($\odot$). 
The region leads to SN Ia explosions is bounded by the solid line. 
\label{zams11ms}}
\end{figure}

     Thus we have obtained the final outcome 
of close binary evolutions for various sets of 
$(M_{\rm WD,0}, M_{\rm MS,0}, a_0)$ 
or $(M_{\rm WD,0}, M_{\rm MS,0}, P_0$).  
Figure \ref{zams11ms} depicts the final outcomes
in the $M_{\rm MS,0} - \log P_0$ plane for $M_{\rm WD,0}= 1.1 M_\odot$.  
Final outcome is either
\def\labelenumi{\theenumi)} 
\begin{enumerate}
\item forming a common envelope (denoted by $\times$) because 
the mass transfer rate at the beginning 
is large enough to form a common envelope, 
i.e., $R_{\rm 1,ph} > a \sim 10 R_\odot$ 
for $\dot M_2 \gtrsim 1 \times 10^{-4} M_\odot$ yr$^{-1}$ 
as seen in Figure \ref{dmdttrvx50z02},
\item triggering an SN Ia explosion 
(denoted by $\oplus$, $\bigcirc$, or $\odot$)
when $M_{\rm 1,WD}= 1.38 M_\odot$,
or 
\item triggering repeated nova cycles (denoted by $\triangle$), i.e., 
$\dot M_2 < \dot M_{\rm low}$ when $M_{\rm 1,WD} < 1.38 M_\odot$.
\end{enumerate}
Among the SN Ia cases, the wind status at the explosion
depends on $\dot M_2$ as follows. 
\begin{enumerate}
\item[2a)] wind continues at the SN Ia explosion for
$ \dot M_{\rm cr} <  \dot M_2 \lesssim 1 \times 10^{-4} M_\odot$ 
yr$^{-1}$ (denoted by $\oplus$).
\item[2b)] wind stops 
before the SN Ia explosion but the mass transfer rate is still 
high enough to keep steady hydrogen shell burning for 
$ \dot M_{\rm st} <  \dot M_2 < \dot M_{\rm cr}$ ($\bigcirc$). 
\item[2c)] wind stops before the SN Ia explosion 
and the mass transfer rate decreases to between 
$ \dot M_{\rm low} <  \dot M_2 < \dot M_{\rm st}$
at the SN Ia explosion ($\odot$).
\end{enumerate}
\par  
     The region producing an SN Ia is bounded by
$M_{\rm MS,0}= 1.8-3.2 M_\odot$ and $P_0= 0.5-5$ d as shown 
by the solid line. 
\def\theenumi{\roman{enumi}} 
\begin{enumerate}
\item The left bound is determined by 
$R^*_{2,f,{\rm CE}} = R_2({\rm ZAMS})$ in Figure \ref{m7m2com2},
where $R_2({\rm ZAMS})$ is 
the minimum radius of the secondary
at the zero age main-sequence (ZAMS).  
\item The right bound corresponds
to the maximum radius at the end of main sequence, after which 
central hydrogen burning vanishes and the star shrinks.
\item The lower bound is determined 
by the decrease in the mass transfer rate
mainly because the secondary mass decreases
to reach the mass ratio $M_2/M_1$ below unity, i.e.,
$\dot M_2 < \dot M_{\rm low}$.
\item The upper bound is limited by the formation 
of a common envelope.
When the mass transfer rate is as high as 
a few times $10^{-5} M_\odot$
yr$^{-1}$ or more, the photosphere of the white dwarf envelope 
reaches the secondary and then swallows 
it, i.e., $R_{\rm 1,ph} \gtrsim a$
(see Fig. \ref{dmdttrvx50z02}).
It may be regarded as the formation of a common envelope.
The binary will undergo a common envelope evolution 
and will not become an SN Ia.
\end{enumerate}
\par
\begin{figure}
\plotone{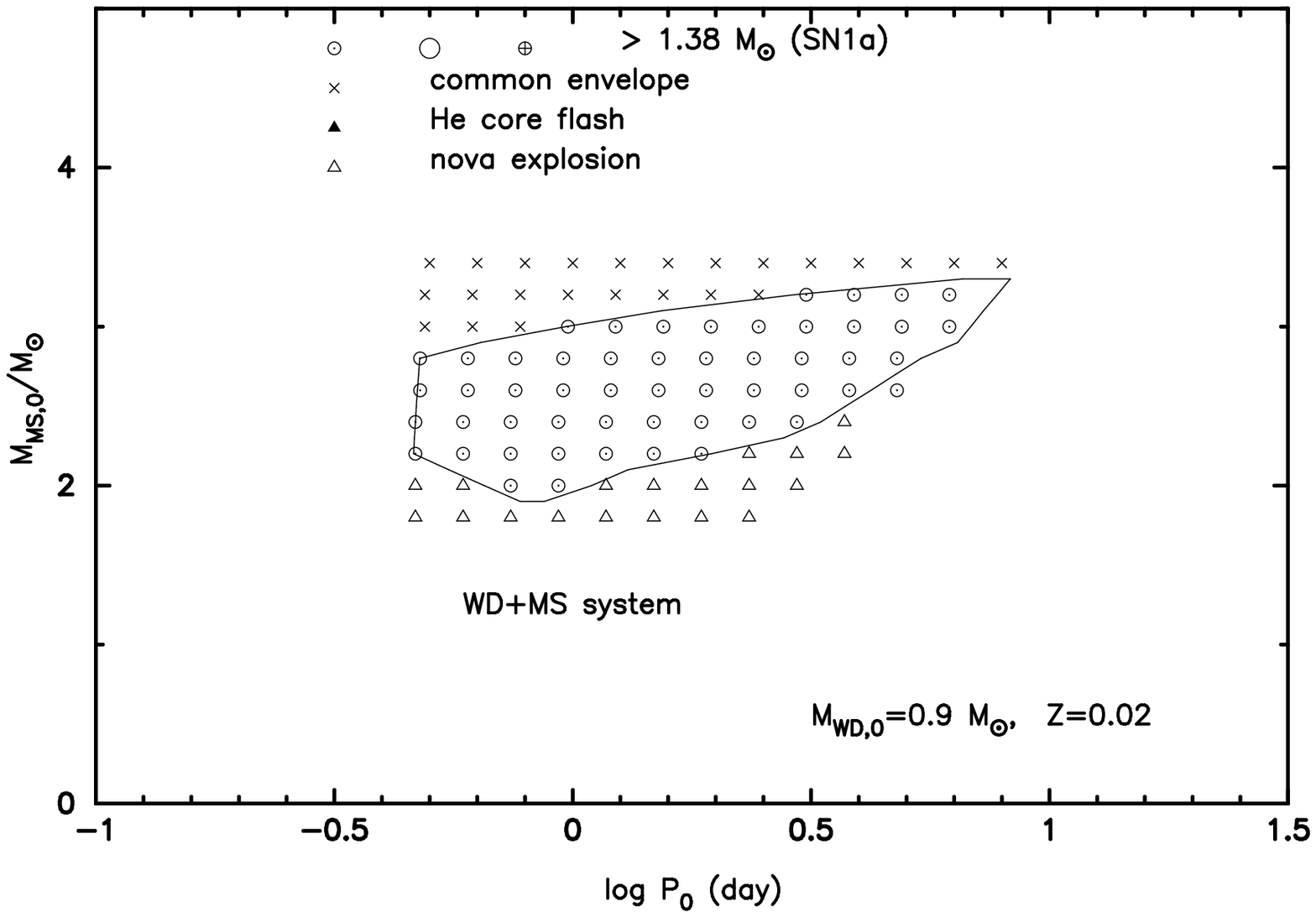}
\caption{
Same as in Fig. 8 but for $M_{\rm WD,0}= 0.9 M_\odot$.  
\label{zams09ms}}
\end{figure}
\begin{figure}
\plotone{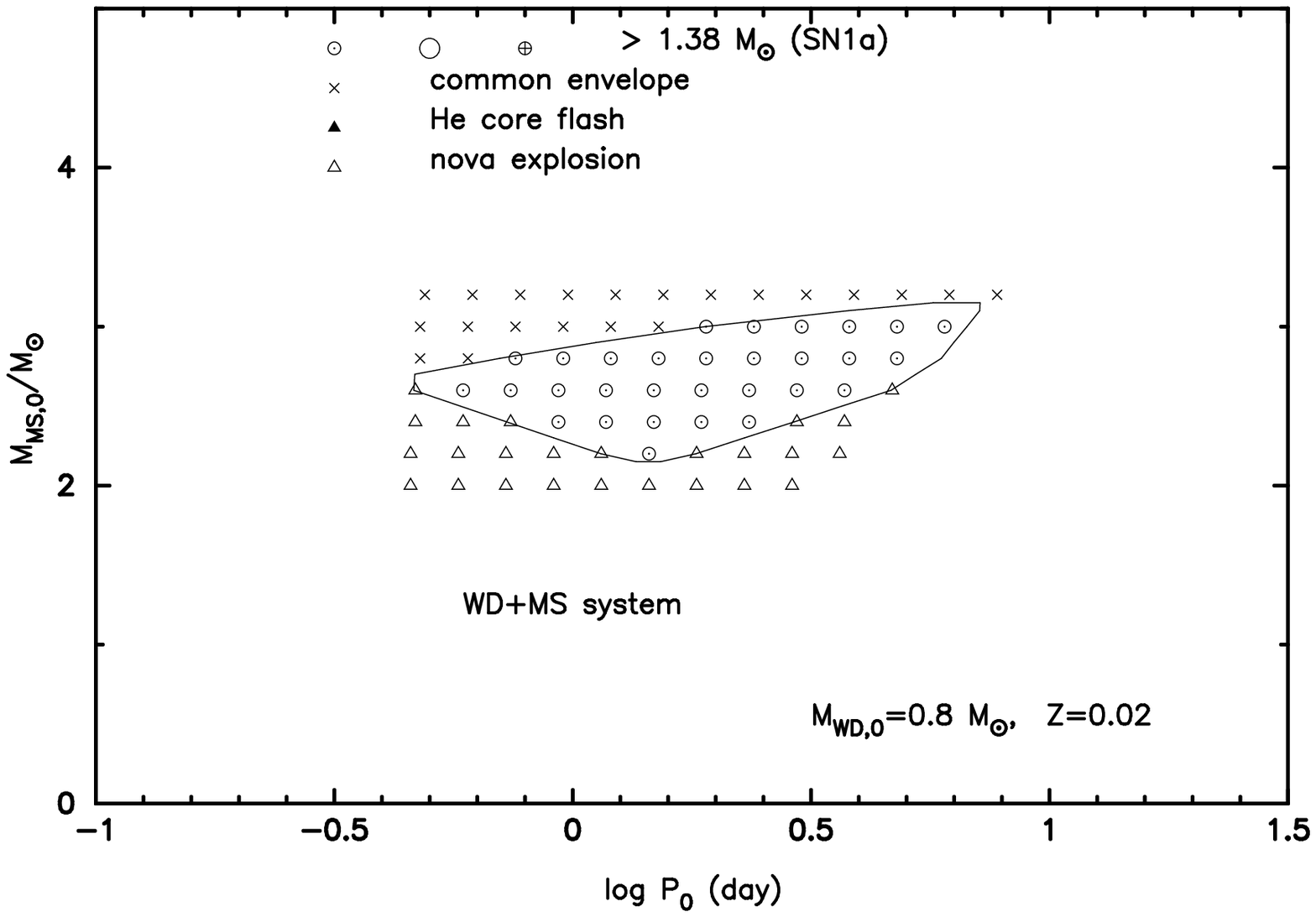}
\caption{
Same as in Fig. 8 but for $M_{\rm WD,0}= 0.8 M_\odot$.  
\label{zams08ms}}
\end{figure}
\begin{figure}
\plotone{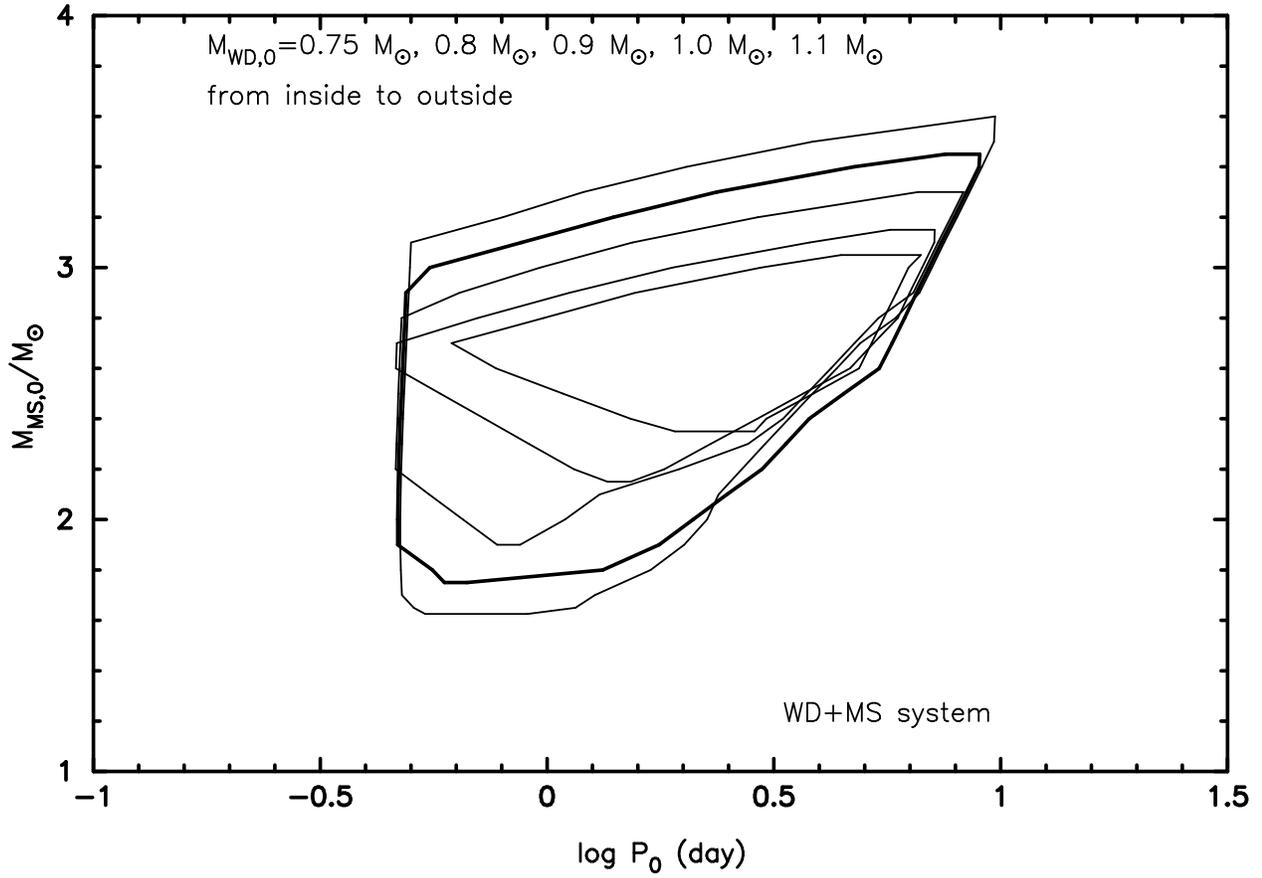}
\caption{
Regions that lead to SN Ia explosions are plotted 
in the $\log P_0 - M_{\rm MS,0}$ plane for five cases of 
the initial white dwarf mass, i.e., $M_{\rm WD,0}= 0.75$,
0.8, 0.9, and $1.1 M_\odot$ (thin solid) 
together with $M_{\rm WD,0}=1.0$ (thick solid).  
The region for $M_{\rm WD,0}= 0.7 M_\odot$ vanishes, however.
\label{zamstot}}
\end{figure}

    The final outcome of the evolutions is also plotted in Figures
\ref{zams10ms}-\ref{zams08ms}
for other initial white dwarf masses of $M_{\rm WD,0}= 1.0 M_\odot$,
$0.9 M_\odot$, and $0.8 M_\odot$, respectively.  
Figure \ref{zamstot} shows the regions that lead to an SN Ia 
for all the white dwarf masses of $M_{\rm WD,0}= 0.75$,
$0.8$, $0.9$, and $1.1 M_\odot$ (thin solid lines)
together with $M_{\rm WD,0}= 1.0 M_\odot$ (thick solid line).
The region for $M_{\rm WD,0}= 0.7 M_\odot$ vanishes.
The shrinking of the upper bound for smaller $M_{\rm WD,0}$ 
is due to larger initial mass ratio 
of $M_{\rm MS,0}/M_{\rm WD,0}$, which enhances the mass transfer rate
at the beginning of mass transfer ({\it stage G}), 
thus resulting in the formation of a common envelope.
The shrinking of the lower bound
can be understood as follows: 
The white dwarf with smaller $M_{\rm WD,0}$ needs to accrete 
more mass from the slightly evolved main-sequence companion.
Thus the companion's mass near the {\it stage I} or {\it J} 
(just before SN Ia explosion) is smaller 
after a considerable part of the hydrogen-rich envelope is transferred
to supply hydrogen-rich matter to the white dwarf.
The thermal time scale of the companion is longer for smaller masses, 
thereby decreasing the mass transfer rate down to the nova region. 
 
\placefigure{zams11ms}
\placefigure{zams10ms}
\placefigure{zams09ms}
\placefigure{zams08ms}
\placefigure{zamstot}

\section{DISCUSSIONS}
     We have estimated the rate of SNe Ia originating from our 
WD+MS systems in our Galaxy 
by using equation (1) of Iben \& Tutukov (1984), i.e.,
\begin{equation}
\nu = 0.2 \cdot \Delta q \cdot 
\int_{M_A}^{M_B} {{d M} \over M^{2.5}} \cdot \Delta \log A 
\quad \mbox{yr}^{-1},
\label{realization_frequence}
\end{equation}
where $\Delta q$ and $\Delta \log A$ denote 
the appropriate ranges of the initial mass ratio 
and the initial separation, 
respectively, and $M_A$, and $M_B$ are the lower and the upper limits 
of the primary mass that leads to SN Ia explosions, respectively. 
For the WD+MS progenitors, we assume that 
$a_{i} \lesssim 1500 ~R_\odot$ in order to obtain 
a relatively compact condition of $a_{f,{\rm CE}}$ after 
the common envelope evolution. 
\par
\begin{figure}
\plotone{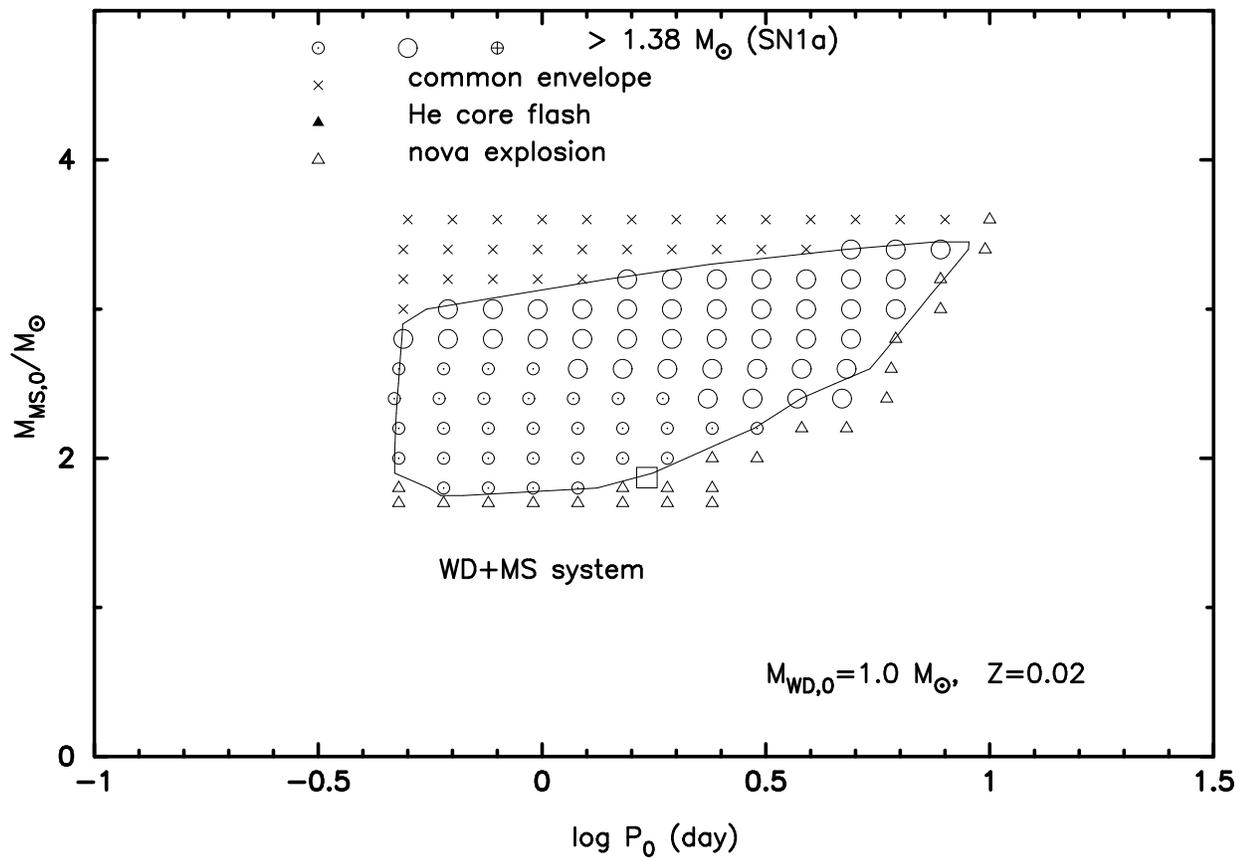}
\caption{
Same as in Fig. 8 but for $M_{\rm WD,0}= 1.0 M_\odot$.  
A box mark ($\Box$) denotes an initial binary parameter of U Sco.
\label{zams10ms}}
\end{figure}

     If the $\sim 1 M_\odot$ C+O WD is a {\it descendant from an AGB star},
its zero-age main sequence mass is $\sim 7 M_\odot$ (see, e.g., 
eq.(11) of \cite{yun95}) 
and the binary separation 
is larger than $a_i \sim 1300 ~R_\odot$ (e.g., \cite{ibe84}).
Its separation shrinks to $a_{f,\rm{CE}} \sim 60 ~R_\odot$ 
after the common envelope evolution for the case of 
$\alpha_{\rm CE} = 1$ and $\sim 2 M_\odot$ secondary.
Then the orbital period becomes $P_0 \sim 30$ d, which is 
too long to become an SN Ia (e.g., \cite{lih97}; see also 
Fig. \ref{zams10ms}). 
Therefore, the WD+MS systems descending from an AGB star 
may be rare as pointed out by Yungelson \& Livio (1998) and 
may not be a main channel to SNe Ia.

\placetable{tbl-1}
\par
     To obtain the realization frequency of our WD+MS system 
{\it descending from a red-giant with a helium core}, 
we have followed total $\sim 500$ evolutions 
with the different initial set of $(M_{1,i},M_{2,i}, a_{i})$
and estimated the appropriate range for the initial separation of 
$\Delta \log A= \log a_{\rm i, max}- \log a_{\rm i, min}$ 
for total $5 \times 5 =25$ cases of $(M_{1,i},M_{2,i})$,
each for the primary mass of $M_{1,0}= 4$, 5, 6, 7, and $9 M_\odot$ 
and the secondary mass of $M_{2,0}= 1$, 1.5, 2, 2.5, and $3 M_\odot$ 
(see Table \ref{tbl-1}).  In Table \ref{tbl-1}, we omit the case of 
$M_{1,0}=4 M_\odot$ because it never leads to SN Ia explosions.
We find that SN Ia explosions occur for 
the ranges of $M_{1,i}= 5.5-8.5 M_\odot$,
$M_{2,i}= 1.8-3.4 M_\odot$, and $\Delta \log A= 0.5$.
We thus obtain the realization frequency of SNe Ia 
from the WD+MS systems as
$\nu_{\rm MS}= 0.0010$ yr$^{-1}$ for $\alpha_{\rm CE}=1$, 
by substituting 
$\Delta q = 3.4/5.5 - 1.8/8.5 = 0.41$, $M_A= 5.5 M_\odot$, 
$M_B= 8.5 M_\odot$, and $\Delta \log A= 0.5$ into 
equation (\ref{realization_frequence}).
For comparison, we have obtained 
a realization frequency of SNe Ia for $\alpha_{\rm CE}= 0.3$.
It is still as high as $\nu_{\rm MS} \sim 0.0007$ yr$^{-1}$,
which is about one fourth of the inferred rate.
Our new rate of $\nu_{\rm MS}= 0.0010$ yr$^{-1}$ 
is about a third of the inferred rate of SNe Ia 
in our Galaxy and much higher than 
$\nu_{\rm MS}= 0.0002$ yr$^{-1}$ (as an upper limit) obtained
by Yungelson \& Livio (1998).  
The reason of their low frequency is probably due to the absence
of the path through the primary's helium star phase
in their scenarios.
However, it should be noted here that Yungelson \& Livio (1998)
have also obtained a realization frequency of
$\nu_{\rm MS} \sim 0.001$ yr$^{-1}$ under the assumption of
no restrictions in their binary evolutions, the conditions of 
which are unlikely.
\par
\begin{figure}
\plotone{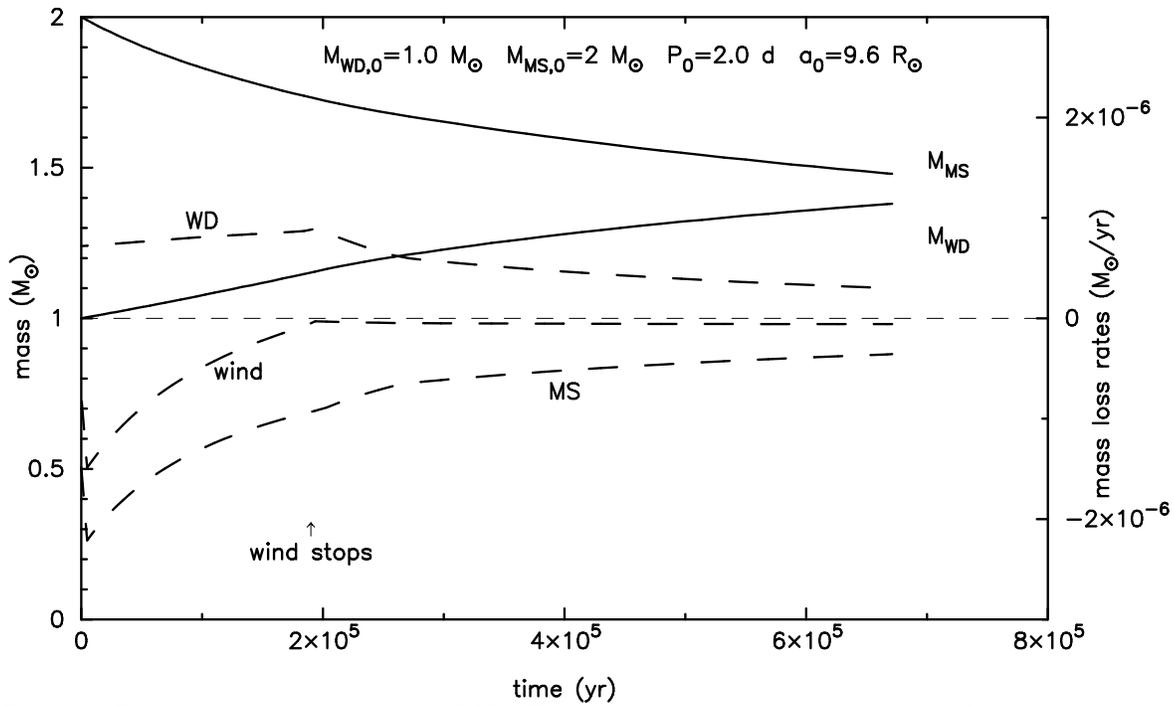}
\caption{
Time evolution of a progenitor of SN Ia.
The white dwarf increases its mass to $1.38 M_\odot$ 
and explodes as an SN Ia.  Solid
lines denote the masses of the white dwarf ($M_{\rm WD}$) 
and the main-sequence companion ($M_{\rm MS}$).
The dashed lines denote, from top to bottom, 
the net mass accretion rate onto the white
dwarf (denoted by ``WD''), 
the wind mass loss rate (denoted by ``wind''), 
and the mass decreasing rate of the companion
(denoted by ``MS''), respectively.
The dashed line shows the average mass loss rate of
$\sim 6 \times 10^{-8} M_\odot$ yr$^{-1}$ due to helium 
shell flashes after ``wind stops.'' 
\label{evlx50z02}}
\end{figure}

A part of
our WD+MS systems are identified as the luminous supersoft X-ray 
sources (SSSs) (\cite{heu92}).  SSSs are characterized by 
a luminosity of $\sim 10^{38}$ erg s$^{-1}$ and a temperature of 
$T \sim 4 \times 10^5$ K $(k T \sim 35$ eV), which have been 
established as a new class of X-ray sources through 
ROSAT observations (e.g., \cite{kah97} for a recent review).
A population synthesis for SSSs has first been done 
by Rappaport, Di Stefano, \& Smith (1994), followed by  
a more complete population synthesis (\cite{yun96}).
These calculations predict the total number of the Galactic SSSs 
of $\sim 1000$ and led to the conclusion that the SSS birth rate is 
roughly consistent with the observation 
(e.g., \cite{kah97} for a review). 
Our SN Ia progenitors should be observed as
an SSS during
the steady hydrogen shell burning phase without 
winds, which is about a few times $10^5$ yr as shown 
in Figure \ref{evlx50z02}.  
Then the number of SSSs from our scenario 
is roughly estimated to be at least 
about $\sim 3 \times 10^5 {\rm ~yr} 
\times \nu_{\rm MS} \approx 300$, which should be added to 1000 by
Yungelson et al. (1996).  The total number is still consistent with 
observations.
\par
     Our WD+MS progenitor model predicts helium-enriched matter 
accretion onto a WD.  Strong He II $\lambda\lambda4686$ lines
are prominent in the luminous supersoft X-ray sources 
(e.g., \cite{kah97} for a recent review) as well as 
in the recurrent novae like U Sco (\cite{han85}; \cite{joh92}) 
and V394 CrA (\cite{sek89}).   Thus
the weakness of the hydrogen emission lines relative to the He II
and CNO lines is very consistent with the requirement 
that the accreted matter and hence 
the envelope of the secondary have a hydrogen-poor (helium-rich)
composition. 
\par
      For SNe Ia, several attempts have been made to detect
signature of circumstellar matter.
There has been no radio detections so far.
Radio observations of SN 1986G have provided the most stringent upper
limit to the circumstellar density as $\dot M / v_{10} = 1 \times
10^{-7} M_\odot$ yr$^{-1}$ (\cite{eck95}), where $v_{10}$ means
$v_{10}= v/10$ km s$^{-1}$.   This is still $10-100$ 
times higher than the density predicted for the white dwarf winds, 
because the WD wind velocity is as fast as $\sim 1000$ km s$^{-1}$.
Further attempts to detect high velocity hydrogen signature are 
encouraged.

\acknowledgments
     We thank the anonymous referee for helpful comments 
to improve the manuscript.
This research has been supported in part by the Grant-in-Aid for
Scientific Research (05242102, 06233101, 08640321, 09640325) 
and COE Research (07CE2002) of the Japanese Ministry of Education, 
Science, Culture, and Sports.

%
%

\begin{deluxetable}{ccccccccc}
\footnotesize
\tablecaption{Initial separations that lead 
to Type Ia supernova explosions ($X=0.50$ and $Z=0.02$). 
\label{tbl-1}}
\tablewidth{0pt}
\tablehead{
\multicolumn{1}{c}{} &
\multicolumn{2}{c}{$5 M_\odot$} &
\multicolumn{2}{c}{$6 M_\odot$} &
\multicolumn{2}{c}{$7 M_\odot$} &
\multicolumn{2}{c}{$9 M_\odot$} \nl
\colhead{} &
\colhead{$\alpha_{\rm CE}=1$} &
\colhead{$0.3$} &
\colhead{$\alpha_{\rm CE}=1$} &
\colhead{$0.3$} &
\colhead{$\alpha_{\rm CE}=1$} &
\colhead{$0.3$} &
\colhead{$\alpha_{\rm CE}=1$} &
\colhead{$0.3$} 
} 
\startdata
$\Delta M_{\rm He}$\tablenotemark{a} & --- & --- & 0.0,0.0 & 0.0,0.2 & 0.1,0.2 & 0.37,0.37 & 0.5,0.6 & 0.6,0.6 \nl
$1.0 M_\odot$ & --- & --- & --- & ---  & --- & --- & --- & --- \nl
$1.5 M_\odot$ & --- & --- & --- & ---  & --- & --- & --- & --- \nl
$2.0 M_\odot$ & --- & --- & 1.9,2.4\tablenotemark{b} & 2.4,2.6 & 1.8,2.3 & 2.3,2.8 & --- & --- \nl
$2.5 M_\odot$ & --- & --- & 1.8,2.4 & 2.3,2.7 & 1.8,2.3 & 2.4,2.8 & 1.8,2.3  & 2.4,2.5 \nl
$3.0 M_\odot$ & --- & 2.3,2.5 & 2.1,2.4 & 2.4,2.7 & 2.0,2.3 & --- & --- & --- \nl
\enddata

\tablenotetext{a}{helium mass transferred to the secondary 
in solar mass units, (minimum, maximum)}
\tablenotetext{b}{initial separation of $\log(a_{\rm i}/R_\odot)$, 
(minimum, maximum)}

\end{deluxetable}

%
%
%
%


%


\begin{thebibliography}{}

%
%
%
%
%
\bibitem[Branch et al. 1995]{bra95}
Branch, D., Livio, M., Yungelson, L.R., Boffi, F.R., \& Baron, E. 1995, 
\pasp, 107, 717

\bibitem[Bressan et al. 1993]{bre93}
Bressan, A., Fagotto, F., Bertelli, G., \& Chiosi, C. 1993, \aaps, 100, 647

%
%
%
%
%
%
%
%
%
%
%
\bibitem[Di Stefano \& Rappaport 1994]{dis94}
Di Stefano, R., \& Rappaport, S. 1994, \apj, 437, 733

%
%
\bibitem[Eck et al. 1995]{eck95}
Eck, C.R., Cowan, J.J., Roberts, D.A., Boffi, F.R., \& Branch, D.
1995, \apjl, 451, L53

\bibitem[Eggleton 1983]{egg83}
Eggleton, P. P. 1983, \apj, 268, 368

%
%
\bibitem[Hachisu \& Kato 1999]{hac99u}
Hachisu, I., \& Kato, M. 1999, \apjl, submitted (U Sco)

\bibitem[Hachisu et al. 1996]{hac96}
Hachisu, I., Kato, M., \& Nomoto, K. 1996, \apjl, 470, L97 (HKN96)

\bibitem[Hachisu et al. 1999]{hac99s}
Hachisu, I., Kato, M., \& Nomoto, K. 1999, \apj, submitted (HKN99)

\bibitem[Hanes 1985]{han85}
Hanes, D. A. 1985, \mnras, 213, 443

\bibitem[Hjellming \& Webbink 1987]{hje87}
Hjellming, M. S., \& Webbink, R. F. 1987, \apj, 318, 794

\bibitem[H\"oflich \& Khokhlov 1996]{hof96}
H\"oflich, P., \& Khokhlov, A. 1996, \apj, 457, 500

%
%
\bibitem[Iben \& Livio 1993]{ibe93}
Iben, I. Jr., \& Livio, M. 1993, \pasp, 105, 1373

\bibitem[Iben \& Tutukov 1984]{ibe84}
Iben, I. Jr., \& Tutukov, A. V. 1984, \apjs, 54, 335

%
%
%
\bibitem[Iglesias \& Rogers 1996]{igl96}
Iglesias, C. A., \& Rogers, F. 1996, \apj, 464, 943

%
%
\bibitem[Johnston \& Kulkarni 1992]{joh92}
Johnston, H. M., \& Kulkarni, S. R. 1992, \apj, 396, 267

%
\bibitem[Kahabka \& van den Heuvel 1997]{kah97}
Kahabka, P., \& van den Heuvel, E. P. J. 1997, \araa, 35, 69

%
%
%
%
%
%
%
%
\bibitem[Kato \& Hachisu 1994]{kat94}
Kato, M., Hachisu, I., 1994, \apj, 437, 802

\bibitem[Kato \& Hachisu 1999]{kat99h}
Kato, M., \& Hachisu, I., 1999, \apjl, 513, in press 
(astro-ph/9901080) 

%
\bibitem[Kato et al. 1989]{kat89}
Kato, M., Saio, H., \& Hachisu, I. 1989, \apj, 340, 509

\bibitem[Kippenhahn \& Meyer-Hofmeister 1977]{kip77}
Kippenhahn, R., \& Meyer-Hofmeister, E. 1977, \aap, 54, 539

\bibitem[Kippenhahn \& Weigert 1967]{kip67}
Kippenhahn, R., \& Weigert, A. 1967, Z. Ap., 65, 251

%
\bibitem[Kovetz \& Prialnik 1994]{kov94}
Kovetz, A., \& Prialnik, D. 1994, \apj, 424, 319

\bibitem[Li \& van den Heuvel 1997]{lih97}
Li, X.-D., \& van den Heuvel, E. P. J. 1997, \aap, 322, L9

%
%
%
%
%
%
\bibitem[Neo et al. 1977]{neo77}
Neo, S., Miyaji, S., Nomoto, K., \& Sugimoto, D. 1977, \pasj, 29, 249

\bibitem[Nomoto 1982a]{nom82a}
Nomoto, K. 1982a, \apj, 253, 798

%
\bibitem[Nomoto 1982b]{nom82b}
Nomoto, K. 1982b, in Supernovae: A Survey of Current Research,
ed. M. Rees \& R. J. Stoneham (Dordrecht: Reidel), 205

\bibitem[Nomoto 1984]{nom84}
Nomoto, K. 1984, \apj, 277, 791

\bibitem[Nomoto \& Hashimoto 1988]{nom88}
Nomoto, K., \& Hashimoto, M. 1988, Phys. Rep., 163, 13

%
\bibitem[Nomoto et al. 1997]{nom97}
Nomoto, K., Iwamoto, K., \& Kishimoto, N. 1997, Science, 276, 1378

%
\bibitem[Nomoto et al. 1994]{nom94}
Nomoto, K., Yamaoka, H., Shigeyama, T., Kumagai, S., \& Tsujimoto, T. 
1994, in Supernovae (Les Houches, Session LIV), ed. S. Bludman et al. 
(Amsterdam: Elsevier Sci. Publ.), 199

%
\bibitem[Nugent et al. 1997]{nug97}
Nugent, P., Baron, E., Branch, D., Fisher, A., \& Hauschildt, P. H. 1997, 
\apj, 485, 812

\bibitem[Paczynski 1971a]{pac71a}
Paczynski, B. 1971a, Acta Astronomica, 21, 1

\bibitem[Paczynski 1971b]{pac71b}
Paczynski, B. 1971b, \araa, 9, 183

%
%
\bibitem[Rappaport et al. 1994]{rap94}
Rappaport, S., Di Stefano, R., \& Smith, J. D. 1994, \apj, 426, 692

%
%
%
%
\bibitem[Saio \& Nomoto 1985]{sai85}
Saio, H., \& Nomoto, K. 1985, \aap, 150, L21  

\bibitem[Saio \& Nomoto 1998]{sai98}
Saio, H., \& Nomoto, K. 1998, \apj, 500, 388

%
%
%
%
%
\bibitem[Sekiguchi et al. 1989]{sek89}
Sekiguchi, K., Catchpole, R. M., Fairall, A. P., Feast, M. W., 
Kilkenny, D., Laney, C. D., Lloyd Evans, T., Marang, F., 
\& Parker, Q. A. 1989, \mnras, 236, 611

%
\bibitem[Segretain et al. 1997]{seg97}
Segretain, L., Chabrier, G., \& Mochkovitch, R. 1997, \apj, 481, 355

%
%
%
%
%
%
%
\bibitem[Tout et al. 1997]{tou97}
Tout, C. A., Aarseth, S. J., Pols, O. R., \& Eggleton, P. P. 1997, 
\mnras, 291, 732


\bibitem[Umeda et al. 1999]{ume99} 
Umeda, H., Nomoto, K., Yamaoka, H., \& Wanajo, S. 1999, \apj, 513, 
in press (astro-ph/9806336)

\bibitem[van den Heuvel et al. 1992]{heu92}
van den Heuvel, E. P. J., Bhattacharya, D., Nomoto, K., \& Rappaport, 
S. 1992, \aap, 262, 97  

\bibitem[Webbink 1984]{web84}
Webbink, R. F. 1984, \apj, 277, 355

%
%
%
\bibitem[Williams et al. 1981]{wil81}
Williams, R. E., Sparks, W. M., Gallagher, J. S., Ney, E. P., 
Starrfield, S. G., \& Truran, J. W. 1981, \apj, 251, 221

%
\bibitem[Yungelson \& Livio 1998]{yun98}
Yungelson, L., \& Livio, M. 1998, \apj, 497, 168

\bibitem[Yungelson et al. 1996]{yun96}
Yungelson, L., Livio, M., Truran, J. W., Tutukov, A., \& Fedorova, A. 
1996, \apj, 466, 890

\bibitem[Yungelson et al. 1995]{yun95}
Yungelson, L., Livio, M., Tutukov, A., \& Kenyon, S. 1995, \apj, 447, 656

\end{thebibliography}
\end{document}